\documentclass[11pt]{article}
\usepackage{graphics}

\begin{document}

\title{ \huge Linear Parabolic Maps on the Torus}

\author{{\Large Karol {\.Z}yczkowski$^{1,2}$   and  Takashi
Nishikawa$^{1}$  }
\vspace{1.0cm}\\
$^1$ Institute for Plasma Research, University of Maryland,\\
College Park, MD 20742, USA\\
\vspace{0.7cm}\\
$^2$ Instytut Fizyki im. M. Smoluchowskiego,\\
 Uniwersytet Jagiello{\'n}ski,
ul. Reymonta 4, 30-059 Krak{\'o}w, Poland }

\maketitle
\vspace{1.2cm}

\centerline{{\small{
e-mail: karol@chaos.if.uj.edu.pl
\ \ \ \ \ \ \ \ \ tnishi@ipr.umd.edu }}}

\vfil

Corresponding author\\
Takashi Nishikawa\\
Institute for Plasma Research, University of Maryland,
College Park, MD 20742, USA\\
TEL +1 301 4051657, FAX +1 301 4051678, tnishi@ipr.umd.edu

\newpage

\begin{abstract}
We investigate linear parabolic maps on the torus. In a generic case
these maps are non-invertible and discontinuous.
Although the metric entropy of these systems is equal to zero, their
dynamics is non-trivial due to folding of the image of the unit square
into the torus. We study the structure of the maximal invariant
set,  and in a generic case we prove the sensitive dependence on the
initial conditions. We study the decay of correlations and
the diffusion in the corresponding system on the plane.
We also demonstrate how the
rationality of the real numbers defining the map
influences the dynamical properties of the system.

\end{abstract}

\vspace{1.2cm}

\centerline{ PACS: {\it 05.45.+b}}
\vspace{0.8cm}


 \section{Introduction}

Linear area-preserving maps on the torus $T^2$ are often analyzed in the
theory of ergodic Hamiltonian systems \cite{CFS82,MM}.
 The map $M$, defined through
\begin{equation}
 \left\{ \begin{array}{lll}
x' & = & ax +by~ |~_{\rm{mod}}~ 1  \\ %\nonumber
y' & = & cx +dy~ |~_{\rm{mod}}~ 1,
\end{array} \right.
\label{map}
\end{equation}
can be represented by the Jacobian matrix
\begin{equation}
 g= \left[ \begin{array}{cc}
    a  & b \\
    c  & d   \end{array} \right].
\label{matrix}
\end{equation}
Due to the area-preserving condition $ad-bc=1$ the matrix $g$
pertains to the non-commutative group $SL(2,{\bf R})$.
Let us emphasize here that the
area-preserving property 
holds for the map when unfolded on the plane. If the dynamics takes place
on the torus, folding causes overlap of some parts of the image of the
unit square, so the area is not conserved.
%This locally compact group is equipped with the Haar measure $\mu$.

For each matrix $g$ there exist its inverse
$g^{-1}=[d, -b; -c, a]$, which defines the inverse map $M^{-1}$ when
the dynamics takes place on the plane $(x,y)$. However, if the dynamics
takes place on the torus $T^2$(modulo $1$ restriction in Eq.~(\ref{map})),
one has to wrap the image of the basic square back into the torus, and
some parts of this image may overlap. Consequently, some points in the
torus may not have any pre-images, and the map (\ref{map}) may not be
invertible on the torus.
%We conjecture that the set $G_{inv}$ of matrices corresponding
%invertible maps on the torus has measure zero with respect to $\mu$.
%Such untypical invertible cases consist for example of systems defined
On the other hand, the systems defined by
matrices with all integer elements are invertible.
The set of these matrices $G_{Int}$
forms a discrete subgroup of $SL(2,{\bf R})$.
Also matrices of the type
$[1, t_1; 0, 1]$ and $[1, 0; t_2, 0]$ represent invertible (but not
continuous) maps on the torus for any non-integer parameters $t_1$ and $t_2$.
These matrices correspond to the horocyclic flows \cite{CFS82} and
form two continuous subgroups of  $SL(2,{\bf R})$,
denoted by $G_{H_1}$ and $G_{H_2}$. 
%Obviously, any product of three
%matrices respectively
%belonging to $G_{Int}$, $G_{H_1}$ and $G_{H_2}$ corresponds to a map
%invertible on the torus.

Dynamical properties of the map (\ref{map}) can be characterized by the
trace $T=$Tr$(g)=a+d$. For $|T|> 2$ the map is {\sl hyperbolic},
eigenvalues of $g$ are real; $z>1$ and $1/z$
(see e.g. \cite{Be}). There exist
infinitely many periodic orbits and all of them are unstable with the
same stability exponent $\lambda_1 =\ln(z)$. The Arnold cat
map defined by the matrix $g=[1, 1; 1, 2]$ is one of the most celebrated examples
of such chaotic dynamical system. Since all elements of $g$ are integer,
this map is invertible on the torus $T^2$(it is an automorphism on the
torus).  For $|T|<2$ the map is called {\sl
elliptic}. The eigenvalues of $g$ are complex and both Lyapunov exponents
are equal to zero. Dynamics of such a system is regular and corresponds
to a rotation.
Properties of the elliptic maps on the torus and some features of the
elliptic-hyperbolic transition were studied by Amadasi and Casartelli \cite{AC}

The intermediate case, $|T|=2$ is called {\sl parabolic}.
Although this case is called by Berry~\cite{Be} 
`special, non-generic, set-of-measure-zero, 
infinitely-improbable-unless-you-deliberately-set-out-to-create-them'
case, we believe that it is worth analyzing, for two reasons.
On one hand, these maps are interesting, as they
lie in between the elliptic and the
hyperbolic
cases, which are very important for several physical 
applications. On the other, the parabolic maps display several
remarkable  properties.
In particular, we show that a generic linear, area-preserving,
parabolic map on the torus displays sensitivity on initial conditions.
Furthermore, we
 demonstrate how the rationality of numbers defining the map
affects the dynamics of the system.
A related study of the systems leading to the interval exchange maps
were discussed in \cite{alex}.

This paper is organized as follows. In section 2, we discuss the
general properties of parabolic maps on the torus and show
corresponding families of 1D maps. In section 3 and 4, we analyze the
case of rational and irrational maps, respectively.
In section 5, we analyze the property  of sensitive dependence 
on the initial conditions. Decay of correlations and the
diffusion rate are analyzed in section 6.

\section{Parabolic maps on the torus}

If the trace $T$ of the matrix (\ref{matrix}) fulfills $|T|=2$, the
linear map on the torus $T^2$ is called {\sl parabolic}.\footnote{
To avoid the confusion let us mention that the 2D maps analyzed in
this work are entirely
different from 1D maps constructed out of parabola, also referred
in the literature as {\sl parabolic maps}.}
The corresponding dynamics is not chaotic and describes a {\sl
shear flow.}
 Since trace of a product of two matrices is usually not
equal to the product of their traces, the set $G_2$ of all
matrices $g$ with the trace equal to 2 and the determinant equal to
unity (or equivalently, the set of all
 area-preserving, linear parabolic maps on the torus)
does not possess a group structure.
Apart from the matrices with integer entries and trace equal to two
(such as $[3, -2; 2, -1]$),
the matrices belonging to the groups $G_{H_1}$ and $G_{H_2}$ also
correspond to
invertible parabolic maps.
If both elements $b$ and $c$ of the matrix (\ref{matrix}) are non-zero,
any matrix $g\in G_2$ may be represented as
\begin{equation}
 g= \left[ \begin{array}{cc}
    1+A  & A/\alpha  \\
    -A \alpha  & 1-A  \end{array} \right],
\label{matr2}
\end{equation}
where both the free parameters $A$ and $\alpha$ are real.
Both eigenvalues of $g$ are equal to unity,
and the eigenvector reads $\bar{v}=(1,-\alpha)$.
For parabolic maps, the two dimensional dynamics on the torus can be
decomposed into a family of 1D maps, which are defined on the lines parallel
to the vector ${\bar{v}}$. The character of the number
$\alpha$  plays therefore a special role: for rational
$\alpha=r/s$, any line parallel to ${\bar{v}}$
forms a closed circle on the torus, consisting of at most $r+s$ lines on
the unit square.
On the other hand, for any irrational value of $\alpha$,
any line parallel to ${\bar{v}}$ winds densely the entire torus.

To reduce the 2D dynamics into a family of 1D maps, let us introduce new
variables $\{p, q\}$ defined as
\begin{equation}
 \left\{ \begin{array}{lll}
p  & =  & x    \\
q  & = & \alpha x + y,
\end{array} \right.
\label{transf}
\end{equation}
where $p \in [0,1)$, and $q \in [0,1+\alpha)$.
In these variables, the map $M$ takes the following form:
\begin{equation}
 \left\{ \begin{array}{lll}
p' & = & p + Aq/\alpha ~ |~_{\rm{mod}}~ 1  \nonumber \\
q' & = & q - [(1-A)q - \alpha p]
- \alpha [p+{A \over \alpha}q],
\end{array} \right.
\label{maptrans}
\end{equation}
where $[p]$ denotes the integer part of the real number $p$.
This formula is derived as following. Consider the line joining
point $(x,y)$ and its image $(x',y')$ in the $xy$-plane. It crosses the horizontal
discontinuities (at integer $y$ values) $[y']$ times, and the vertical 
ones $[x']$ times. Then the above formula is obtained from 
$q' = q - [y'] - \alpha [x']$
since the slope of the line is $-\alpha$.
Note that if $q_0 = \alpha x_0 + y_0$ is the initial $q$ value, 
the variable $q$ can be represented as $q = q_0 + l_1 + l_2 \alpha$
with integers $l_1$ and $l_2$, and the dynamics of $q$ can be regarded as
a motion on the integer lattice $(l_1,l_2)$.
For some purposes, it is convenient to ignore the operation modulo 1 
in Eq. (\ref{maptrans}), which corresponds to considering the dynamics on
the 
$xy$-plane instead of the torus. The dynamics is described by the family of
1D maps
\begin{equation}
P'  =  P + {A \over \alpha} \left\{ q_0 - \alpha [P] - [q_0 - \alpha P]\right\},
\label{tran2}
\end{equation}
where $P \in {\bf R}$ and the initial condition $\alpha x_0 + y_0 = 
q_0 \in [0, 1+\alpha)$ is treated as a parameter.
The slope of the above
map is constant and equal to unity. Thus the Lyapunov exponent is equal
to zero, in agreement with the assumed parabolicity of the 2D map
$M$. For rational $\alpha=r/s$, the map (\ref{tran2}) is periodic with
period $s$. In the simple case $\alpha=1$, the map (\ref{tran2})
becomes a {\sl lift of degree one}, often analyzed in the theory
of dynamical systems (see e.g. \cite{katok}).

\section{Rational parabolic maps on the torus}

\begin{figure}
\resizebox{\textwidth}{!}{\includegraphics{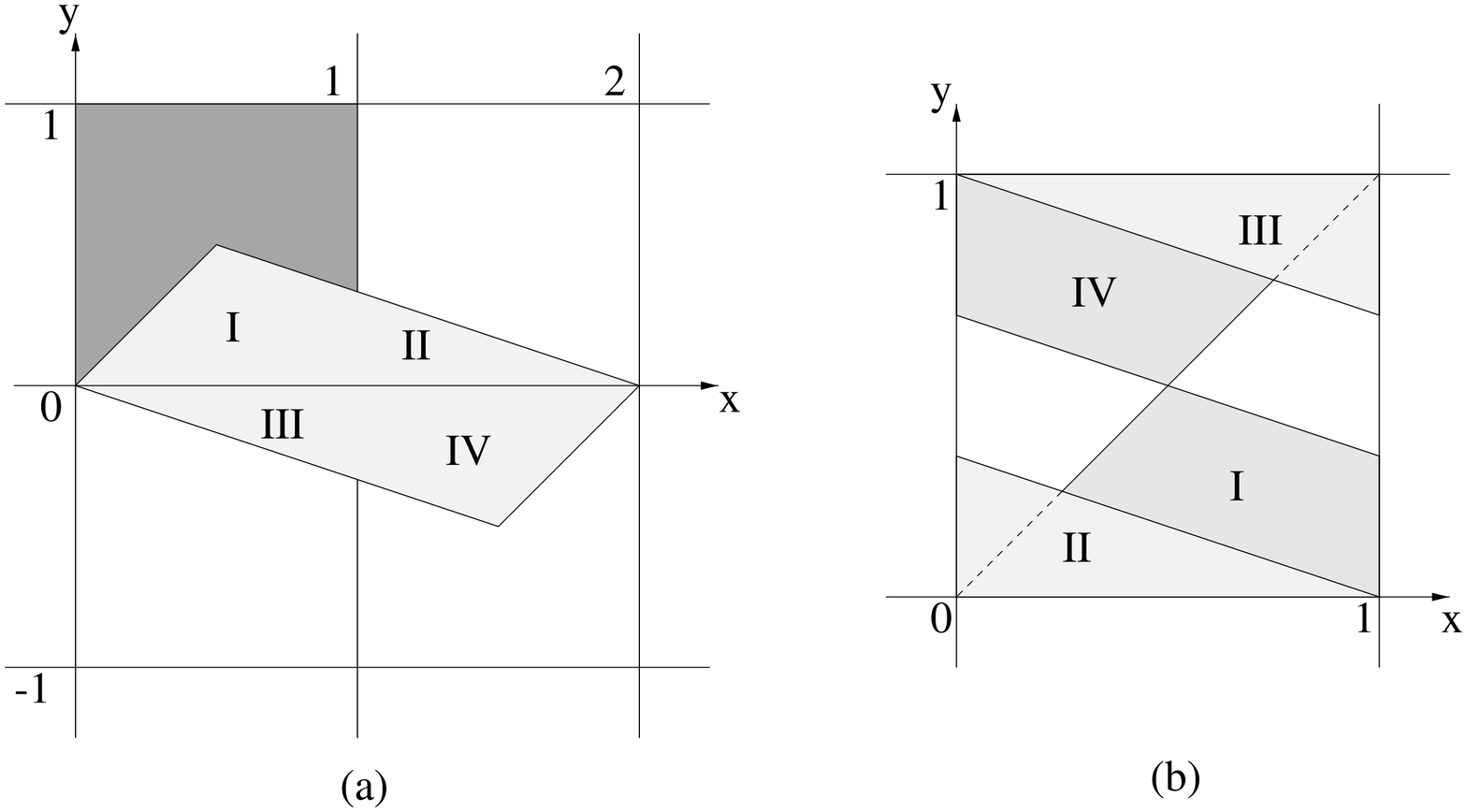}}
\caption {First iterate of the parabolic symmetric map $M_s$ (a) on
the plane; (b) on the torus.(rescaled size)}
\label{fig:sym1}
\end{figure}

We shall call a parabolic map on the torus {\sl rational},
if the parameter $\alpha$, which enters the matrix $g$ given by
(\ref{matr2}), is rational.   If $\alpha$, $A$ and $A/\alpha$ are
rational but not integer, the corresponding map $M$ is discontinuous
on all sides of the square and is not invertible on the torus. A simple
example is given by $\alpha=1$ and $A=1/2$, which represents
the map $M_s$
\begin{equation}
 \left\{ \begin{array}{lll}
x' & = & ~{3 \over 2} x +{1 \over 2} y~ |~_{\rm{mod}}~ 1  \\ %\nonumber
y' & = & -{1 \over 2} x +{1 \over 2} y~ |~_{\rm{mod}}~ 1,
\end{array} \right.
\label{maps}
\end{equation}
associated to the matrix $g_s:=[3/2, 1/2; -1/2, 1/2]$.
Figure~\ref{fig:sym1}(a) represents
the first iterate of the basic square on the plane.
Due to the wrapping of this figure back into the torus, some parts of it
overlap, and some fragments of the torus
(denoted in white in Fig.~\ref{fig:sym1}(b) do not have any pre-images with respect
to $M_s$. Subsequent images of this set are spread over entire torus,
which is what makes the dynamics of the system interesting.

\begin{figure}
\setlength{\unitlength}{1in}
\begin{picture}(6,6)(0,0)
\put(0.5,0.5){\resizebox{5in}{!}{\includegraphics{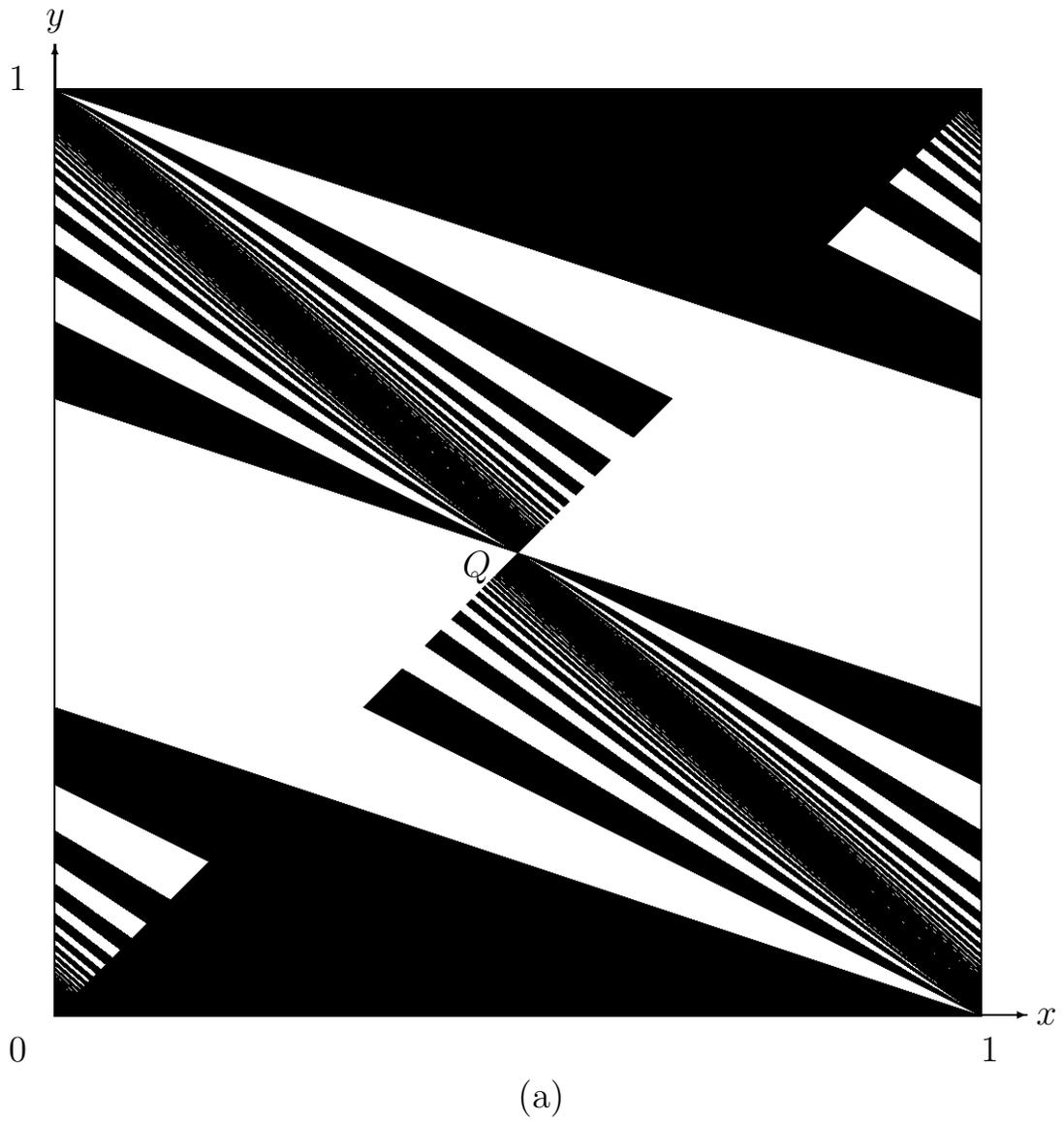}}}
\put(0.25,0.25){\Large{0}}
\put(0.25,5.5){\Large{1}}
\put(5.5,0.25){\Large{1}}
\put(0.5,0.5){\framebox(5,5)}
\put(0.5,5.5){\vector(0,1){0.25}}
\put(5.5,0.5){\vector(1,0){0.25}}
\put(0.45,5.85){\Large{$y$}}
\put(5.8,0.45){\Large{$x$}}
\put(3,0){\Large{(a)}}
\put(2.7,2.87){\Large{$Q$}}
\end{picture}
\caption {(a) Invariant set $S$ of the symmetric parabolic
map $M_s$;(b) magnification of $S$ in the vicinity of the critical point
$Q$.}
\label{fig:sym2}
\end{figure}

Since $\alpha=1$, any trajectory originating from a single point
is restricted to a circle on the torus represented by at most two parallel
lines (of the same slope equal to $-1$) in the unit square.
Therefore the map $M_s$ does not have an
attractor in the sense of attracting single trajectories. 
Figure~\ref{fig:sym2} presents a graph obtained from
ten thousand iterations of
trajectories originating from ten thousand different initial points
 randomly chosen
according to the uniform distribution on the torus. To avoid transient
effects, first hundred points of each trajectory are not marked.
Denote the maximal invariant set \cite{papc}
(black in the figures) by
\begin{equation}
S=\bigcap_{n\in{\bf N}} M^n(T^2).
\end{equation}
This set forms a support of a `natural' invariant measure $\mu_*$ of
the map $M$, which may be generated by
iterating the initially uniform
(Lebesgue) measure by the associated Frobenius-Perron operator.
The measure $\mu_*$ need not cover the set $S$ uniformly.

\begin{figure}
\hspace{1in}
\begin{center}
\resizebox{5in}{!}{\includegraphics{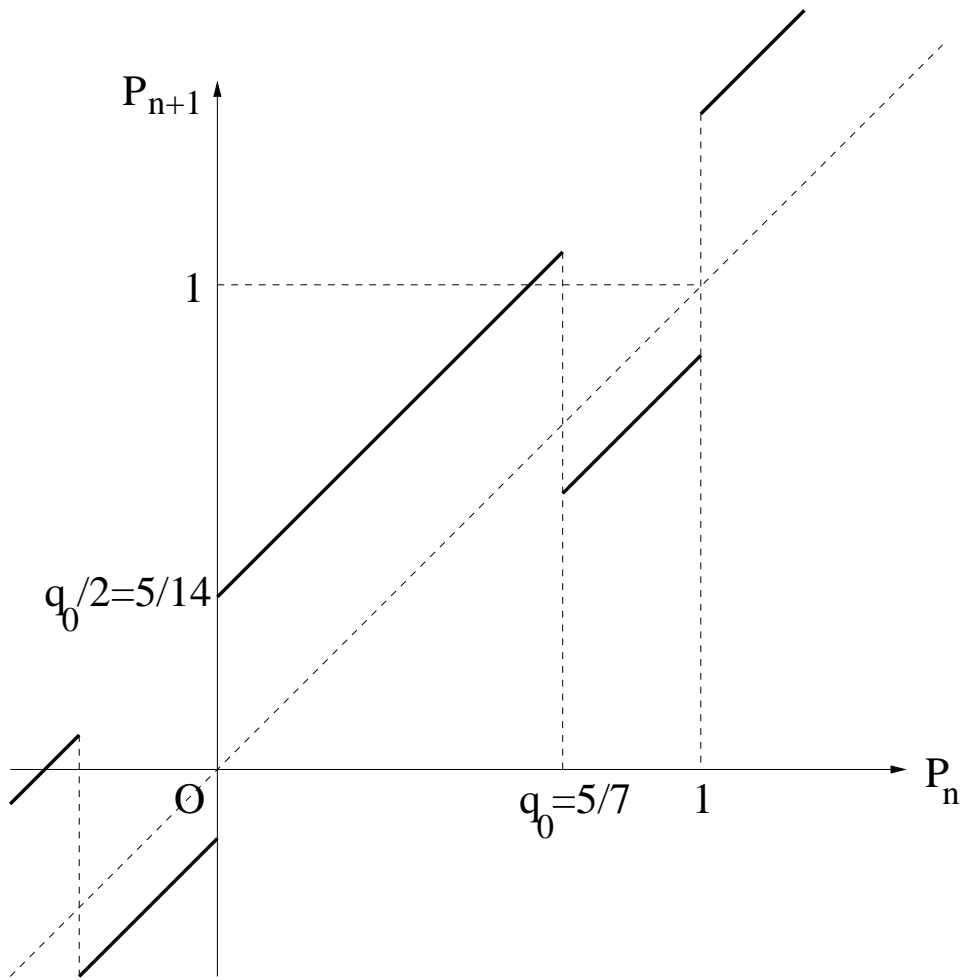}}
\end{center}
\caption{Periodic 1D map derived from the symmetric map $M_s$ for $q=5/7$.}
\label{fig:1Dmap}
\end{figure}

{\bf Remark 1.} Observe the rotational symmetry of $S$ around the
central point $Q=(1/2,1/2)$. Moreover, for any fragment of the set $S$
(black in the figure), one can find the corresponding white fragment
belonging to ${\bar{S}}=T^2\setminus S$. For example, the two large
black fragments at the top and the bottom of the square  form a
parallelogram on the torus, which corresponds to the white
parallelogram consisting of the two large white fragments on the left and
on the right of the unit square.
 Thus there exists a symmetry in shifting the figure by
vector $(1/2, 1/2)$ and interchanging the colors. Therefore, the
volume of the invariant set $S$ is equal to $1/2$. In other words,
for $\alpha=1, A=1/2$,
the invariant set of the 1D map (\ref{tran2})   (see Fig.~\ref{fig:1Dmap})
has volume $1/2$ independently of the parameter $q$.

The fact that the invariant set $S$ contains a ball of a positive radius
implies
that its dimension is equal to 2.
Figure~\ref{fig:sym2}(b) presents the magnification of the invariant set in the
vicinity of the point $Q$. Note an infinite sequence of interlacing
black and white parts of the figure which accumulate on the diagonal
line and form some kind of self-similar structure.

{\bf Remark 2.} Numerical experiments suggest that the invariant set $S$
is structurally stable.
 It is thus  possible to get a practical approximation of the
set $S$ by iterating a single trajectory of an modified
system close to the analyzed map $M_s$. Perturbing any of four elements
of $g_s$  (say element $a$ by putting $a \to a+\varepsilon$
with a real parameter $0<\varepsilon << 1$),
we get a perturbed map $M_{\varepsilon}$.
Formally, this matrix does not belong to $G_2$, since the area-preserving
property and parabolicity, $T=2$, are $\varepsilon$-violated. The perturbation
 couples
all 1D maps together, so that by iterating one single trajectory we
obtain the picture of the invariant set $S_{\varepsilon}$ of the map
$M_{\varepsilon}$. The numerical results show that this set approximates the
set $S$ in the Hausdorff metric, while the quality of this approximation improves
when decreasing the perturbation strength $\varepsilon$.

\section{Irrational parabolic maps on the torus}

\begin{figure}
\setlength{\unitlength}{1in}
\begin{picture}(6,6)(0,0)
\put(0.5,0.5){\resizebox{5in}{!}{\includegraphics{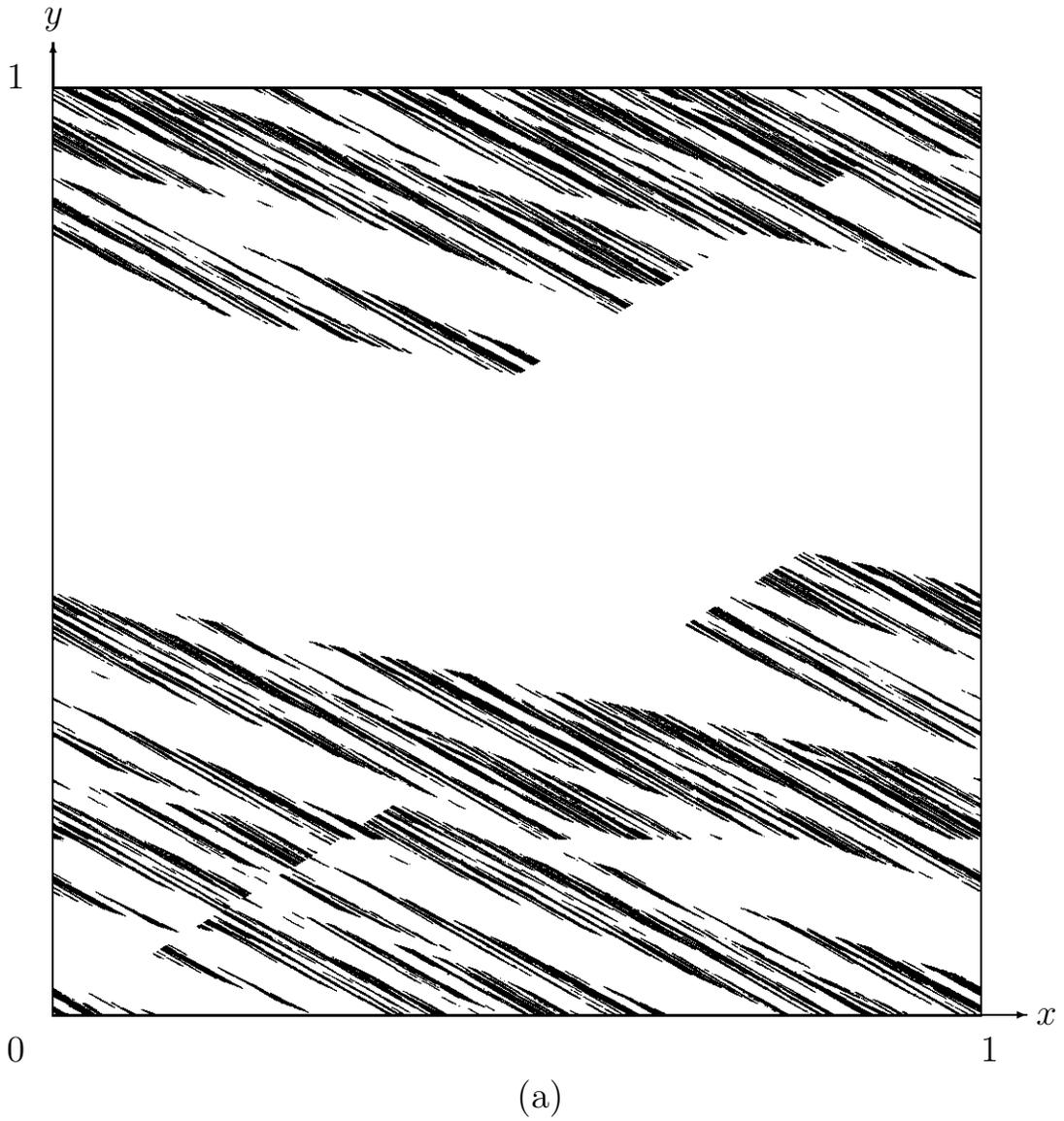}}}
\put(0.25,0.25){\Large{0}}
\put(0.25,5.5){\Large{1}}
\put(5.5,0.25){\Large{1}}
\put(0.5,0.5){\framebox(5,5)}
\put(0.5,5.5){\vector(0,1){0.25}}
\put(5.5,0.5){\vector(1,0){0.25}}
\put(0.45,5.85){\Large{$y$}}
\put(5.8,0.45){\Large{$x$}}
\put(3,0){\Large{(a)}}
\end{picture}
\caption{The maximal invariant set $S_\gamma$ for the irrational parabolic map
$M_{\gamma}$ defined by  $\alpha= \gamma$ (a);
and the rational maps related to the Fibonacci sequence
$\alpha=5/8$ (b); and  $\alpha=2/3$ (c).}
\label{fig:gold}
\end{figure}

A generic matrix belonging to $G_2$ contains irrational elements.
To analyze such a case in some details we take the
golden mean $\gamma:=(\sqrt{5}-1)/2$  for the parameter $\alpha$
entering the matrix (\ref{matr2}), and set $A=1/2$ arbitrarily. This
choice corresponds to the map $M_{\gamma}$ given by
\begin{equation}
 \left\{ \begin{array}{lll}
x' & = &  \ ~{3 \over 2} x + {1 \over 2\gamma} y~ |~_{\rm{mod}}~ 1  \\
%\nonumber
y' & = & -{\gamma \over 2} x + {1 \over 2} y~ |~_{\rm{mod}}~ 1.
\end{array} \right.
\label{mapgg}
\end{equation}
Due to the irrationality of the parameter $\alpha$,
the line parallel to the eigenvector of $g_{\gamma}$ winds densely around the
entire torus. In other words, the 1D map (\ref{tran2}) is not periodic.
Thus, by iterating (almost every) single trajectory by the
corresponding map $M_{\gamma}$ given by Eq. (\ref{map}),
we obtain full information concerning the invariant attracting set
$S_{\gamma}$. Such a set is shown in Fig.~\ref{fig:gold}(a).
Numerical analysis allows us to estimate
the capacity dimension of this set  $D_0\approx 1.71\pm 0.05$ and the
correlation
dimension $D_2\approx 1.70\pm 0.02$. These data suggest strongly a
fractal structure of the set $S_{\gamma}$.
Note that the transformation $M_{\gamma}$ is volume-preserving on the
plane, and the existence of the attractor for this map is solely due to
the  modulo $1$ condition, which corresponds to
wrapping the image of the unit square back on the torus.

Every irrational number can be approximated by rational numbers by
truncating its expansion into the continuous fraction. For the golden
mean, the consecutive approximations are given by the ratios of the
adjacent Fibonacci numbers. Taking such approximations as
$\gamma_1=2/3$ and $\gamma_2=5/8$, we get two rational maps
(\ref{matr2}) defined by $\alpha_1=2/3,A_1=1/2$ and $\alpha_2=5/8,A_2=1/2$.
%%%% small additions and changes !!!!
Estimation of the volume of the corresponding invariant sets $S_1$ and
$S_2$ gives $V_1=0.3$ and $V_2=0.2$, respectively. The volume decreases
for rational maps obtained for better rational approximations of $\gamma$
characterized by larger denominators. These results suggest that the
dimensions of the sets $S_1$ and $S_2$ are equal to $2$.
However, an unusually slow convergence of the standard numerical
procedure made it difficult to get a reliable numerical estimations 
of these dimensions.
Observe similarities between $S_1$ and $S_2$ (depicted in
Fig.~\ref{fig:gold}(c) and (b))
and the invariant set $S_{\gamma}$ of the irrational map presented in
Fig.~\ref{fig:gold}(a). Even though the sets $S_2$ and $S_{\gamma}$ look similar, we
conjecture that their structure is very different.
Invariant set $S_2$, corresponding to the rational map, is conjectured
to have a non-zero volume and the capacity dimension equal to $2$. This
contrasts the properties of the set $S_{\gamma}$
invariant under the irrational map $M_{\gamma}$.

\section{Discontinuity and sensitivity on initial conditions}

Let us denote by $D$ the boundary of the unit square.
The linear, parabolic linear area-preserving map (\ref{map})
is discontinuous at $D$ with probability one
 with respect to the Haar measure on $SL(2,{\bf R})$ (the only case that it is not
discontinuous is when all the entries of the matrix are integer).
 The discontinuity can be measured
by $\Delta$, the minimal size of the
distances d$(M(0,y),M(1,y))$ and d$(M(x,0),M(x,1))$ on the torus $T^2$,
which depends only on the parameters of a parabolic map.
It follows from the definition of $\Delta$ that the condition $\Delta = 0$
is satisfied only by countable number of pairs $(A,\alpha)$. Thus, a generic
parabolic map is characterized by a positive $\Delta$.
 The lines of discontinuity have
implications for the dynamics of the system, since
they split a bundle of neighboring trajectories. Similar effects were
reported for elliptic maps on the torus in \cite{As96,As97}.

Let us consider the set of all pre-images of the set $D$.
 Although the inverse map
$M^{-1}$ may not exist, we define the pre-images
$M^{-k}(D)$ by
\begin{equation}
M^{-k} (D)  = \{ (x,y) : M^k(x,y)\in D {\rm{~but~}}
  M^n(x,y) \notin D {\rm{~for~}} 0\le n < k\}.
\label{pre}
\end{equation}
In an analogous way, $M^k(D)$ denotes the $k$-th image of $D$ with
respect to $M$. Moreover, following \cite{As96}, we define
\begin{equation}
  D^{-}  = \bigcup_{k\in N} M^{-k} (D) {\rm{~and~}}
  D^{\pm}  = \bigcup_{k\in Z} M^{k} (D).
\label{premul}
\end{equation}
Since $D^-$ is a countable union of sets of finite one-dimensional Lebesgue measure,
it has two-dimensional Lebesgue measure zero.
However, its closure
$\overline{D^-}$ may have a positive measure.
Indeed that is the case for our parabolic map,
 and the following proposition is true.

{\bf Proposition 1.} {\sl For a
parabolic map $M$ with discontinuity,
the set  $D^-$ is dense in the torus $T^2$ and its closure
${\overline{D^-}}$ has the full measure.}

{\bf Proof.} Let $B_{x,\varepsilon}$ be
 the $\varepsilon$-ball around any point $x$ on the torus.
We have to show that it must intersect with $D^{-}$.
Suppose that this $\varepsilon$-ball and all of its images
 $M^n(B_{x,\varepsilon})$ do not intersect with $D$.
Due to the shearing effect of the map the diameter of
$M^n(B_{x,\varepsilon})$ grows with $n$ as
\begin{equation}
 \mbox{diam}(M^n(B_{x,\varepsilon})) = 2\varepsilon\sqrt
{n^2 A^2 \left(\alpha + {1 \over \alpha} \right)^2 + 1}.  
\end{equation}
Therefore if $n$ is large enough, the diameter will be larger than $\sqrt{2}$
 and the image has to intersect the boundary $D$. %$\Box$

\begin{figure}
\setlength{\unitlength}{1in}
\begin{picture}(6,6)(0,0)
\put(0.5,0.5){\resizebox{5in}{!}{\includegraphics{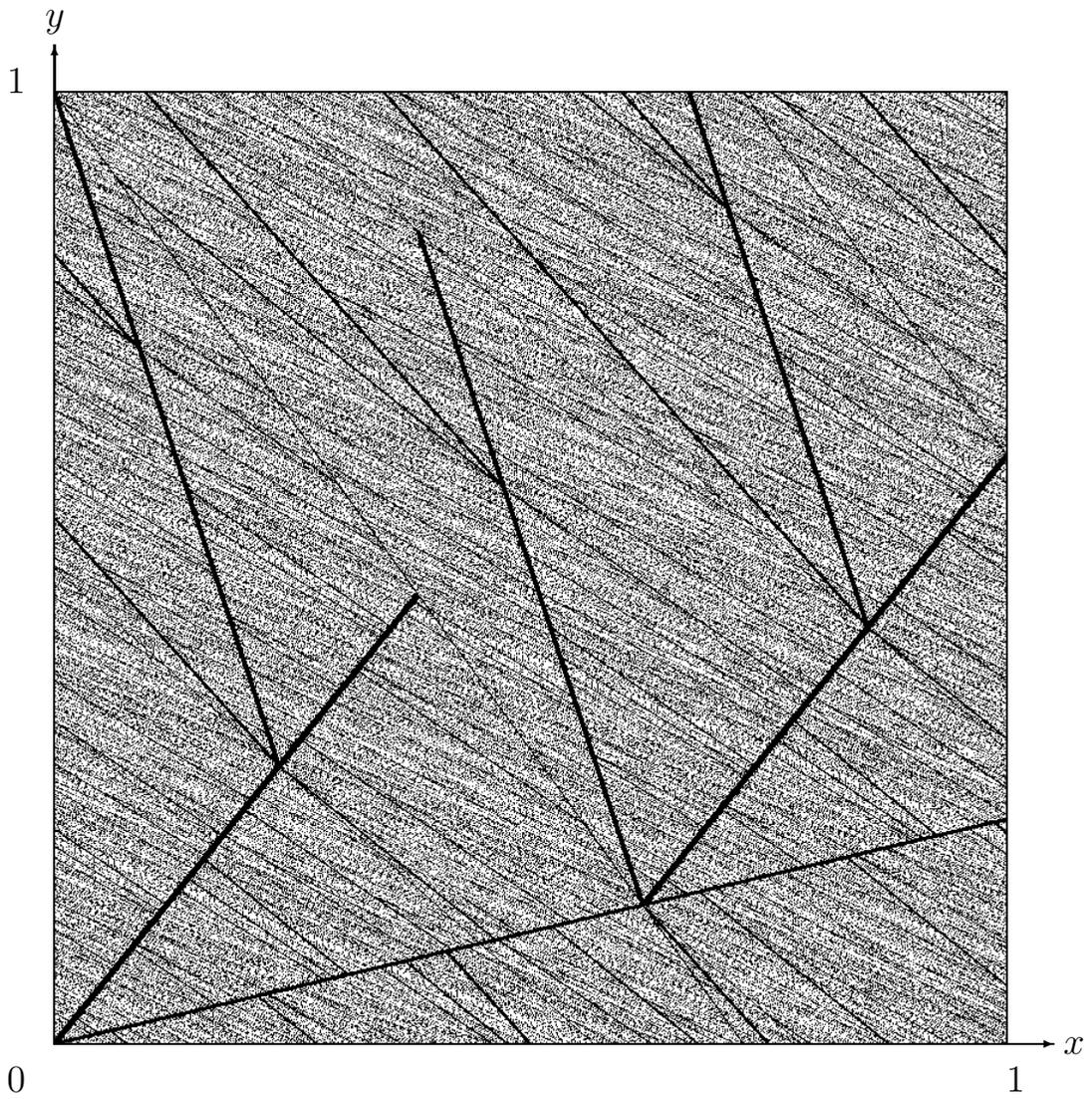}}}
\put(0.25,0.25){\Large{0}}
\put(0.25,5.5){\Large{1}}
\put(5.5,0.25){\Large{1}}
\put(0.5,0.5){\framebox(5,5)}
\put(0.5,5.5){\vector(0,1){0.25}}
\put(5.5,0.5){\vector(1,0){0.25}}
\put(0.45,5.85){\Large{$y$}}
\put(5.8,0.45){\Large{$x$}}
\end{picture}
\caption{The set $D^-$ of the pre-images of the discontinuity lines up to $k=50$
iterates for the system $M_{\gamma^2}$.
Thick lines correspond to the pre-images $M_{\gamma^2}^{-k}(D)$ for low values
of $k$, but the thickness is merely
an artifact of the numerical procedure.
The picture is produced using the $800 \times 800$ mesh.}
\label{fig:preim}
\end{figure}

Figure~\ref{fig:preim} shows the set $D^{-}$ for the
irrational map $M_{\gamma^2}$ defined by $\alpha=A=\gamma^2$
computed by a method similar to that used in \cite{As97},
(the map $M_{\gamma^2}$ is distinguished by the fact that one element of
the matrix $g$ is equal to $\gamma$).
Proposition 1  allows us to formulate main result of the present paper
concerning dynamics of a generic parabolic map.

{\bf Corollary.} {\sl The dynamics generated
 by a generic parabolic map $M$
 has {\it sensitive
dependence} on initial conditions in the torus $T^2$}.

This property is understood as in the study of Ashwin \cite{As96}
regarding elliptic discontinuous maps on the torus.

 {\sl Sensitive dependence on
initial conditions} in a set $\Omega$ means that for all $x_1\in \Omega$
there
exists an $\epsilon>0$ such that for all $\delta>0$ there exists a point
$x_2\in \Omega$ with $|x_1-x_2|< \delta $  and
$|M^n(x_1)-M^n(x_2)|>\epsilon $ for some $n$.

{\bf Proof.} As in the proof of the Proposition 1, any
$\varepsilon$-ball will intersect $D$
after some iterations. Then the next iteration will split
the ball into two disjoint sets separated by $\Delta$,
which is positive for a generic map $M$.
Hence it has sensitivity on the initial conditions. %$\Box$

Note that the proofs do not require that the map be irrational.
 It seems, however, that in the case of irrational maps,
there is also another mechanism for the sensitivity due to the
 motion on the lines parallel to the eigenvector,
whereas the rational maps do not have sensitivity along the eigenvector.
Moreover, the above reasoning shows that the discontinuous parabolic
maps fulfil a slightly different property of {\sl sensitive dependence}
introduced in the paper of Guckenheimer \cite{Gu79}.
 In any case, we emphasize that for this parabolic systems the
sensitivity on initial conditions is not related with the
Lyapunov exponents (which are zero in this case), but due to the
discontinuity of the map.

\section{Diffusion and the decay of correlations}

Decay of correlations for a family of linear invertible
parabolic maps on the torus was recently analyzed by Courbage and
Hamdan \cite{CH97}. Although these systems are not chaotic
(zero Kolmogorov-Sinai entropy), the correlations decay fast.
 While systems with
sub-exponential correlations decay are found to be generic,
a class of systems with exponential decay rate was found.

\begin{figure}
\setlength{\unitlength}{1in}
\resizebox{\textwidth}{!}{
\begin{picture}(7,8.5)(0,0)
\put(3.6,8.1){\scalebox{1.5}{(a)}}
\put(0.55,6.1){\rotatebox{90}{\scalebox{1.5}{$C_q(n)$}}}
\put(3.6,4.3){\scalebox{1.5}{n}}
\put(0.5,4.5){\resizebox{6in}{3.5in}{\includegraphics{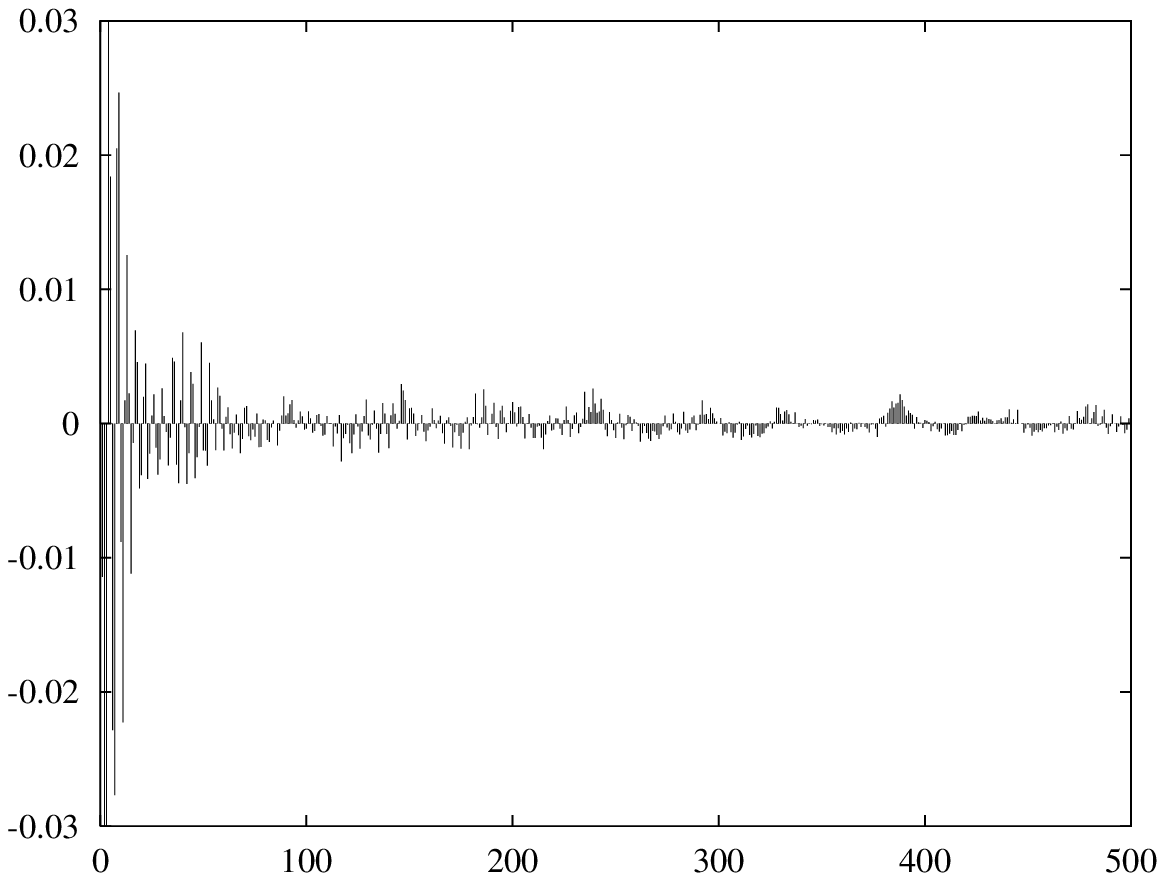}}}
\put(3.5,3.8){\scalebox{1.5}{(b)}}
\put(0.35,1.7){\rotatebox{90}{\scalebox{1.5}{$\hat{S}(\omega )$}}}
\put(3.6,0.1){\scalebox{1.5}{$\omega$}}
\put(0.5,0.2){\resizebox{6in}{3.5in}{
\rotatebox{0}{\includegraphics{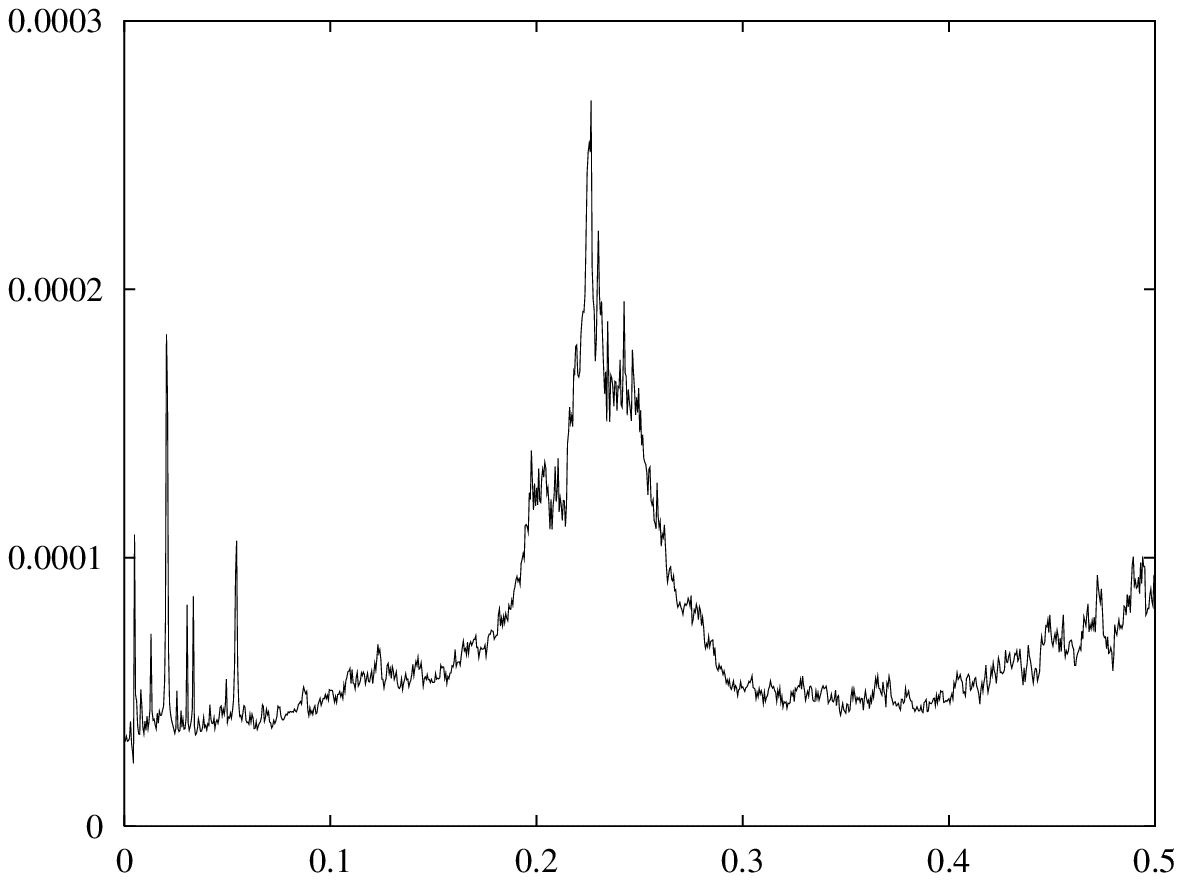}}}}
\end{picture}
}
\caption{(a) Force-force correlation function $C_q(n)$
 for the system $M_{\gamma^2}$ (b) and its Fourier transform $\hat{S}(\omega )$.}
\label{fig:ac}
\end{figure}

Analogously, the non-invertible parabolic maps, discussed throughout
this paper, display fast decay of correlations, for generic,
irrational maps.
%Sensitive dependence on initial conditions manifests itself also in the
%decay of correlations.
Figure~\ref{fig:ac}(a) presents the normalized
$q$--$q$ autocorrelation function
\begin{equation}
  C_q(n)= \frac{ \langle (q_0 - \langle q_0\rangle )(q_n -\langle q_n\rangle)\rangle}
	{ \langle ( q_0 - \langle q_0\rangle )^2 \rangle },
\label{corr1}
\end{equation}
where the average is taken over initial points with
 respect to the natural  measure. Numerical data
are obtained from an ensemble of $10^6$ initial points chosen randomly according to the
natural measure $\mu_*$ for the irrational map $M_{\gamma^2}$.
The fast decay of correlations, visible in this plot, was also
 observed for several other
initial conditions. This suggests that the autocorrelation function
is independent of the initial
conditions, which supports the conjecture
that the dynamics is ergodic with respect to the invariant measure
$\mu_*$.  As before, the fast correlation decay cannot be
attributed to the positive Lyapunov exponent, but rather to
discontinuity of the map $M_{\gamma^2}$.

Figure~\ref{fig:ac}(b) shows the Fourier transform of the autocorrelation function in (a),
which is equal to the power spectrum of the $q$--time series. Notice the broad band continuous
component in the spectrum similar to what is normally seen for chaotic maps.
Also notice that there are several peaks, 
suggesting some periodic components both in $q$--time series
and the $q$--$q$ autocorrelation function.

In order to discuss the diffusion process it is convenient
to work with the unfolded dynamics on the plane
(i.e. the operation modulo $1$ in Eq. (\ref{maptrans}) is not performed
and the variable $P$ in eqn.(\ref{tran2}) takes arbitrary real values).
Diffusion coefficient may be defined as
\begin{eqnarray}
\nonumber
D_p &=& \lim_{n \to \infty} D_p(n)\\
 &=&  \lim_{n \to \infty} {
 \langle (P_n - \langle P_n \rangle)^2 \rangle \over n} .
\label{diff2}
\end{eqnarray}
Since $P_{n+1}-P_n=(A/\alpha) q_n$ the position--position correlation
function $C_q$ is equal to $A/\alpha$ times the force--force
correlation function in the direction $p$.
This function determines the diffusion rate $D_p$ along the $p$ axis.
An argument similar to one in \cite{meiss} shows
\begin{eqnarray}
\nonumber
  D_p &=& \lim_{n \to \infty}{1 \over n} \sum_{i,j=1}^n
	\langle (\Delta p_i - \langle \Delta p_i \rangle)
	(\Delta p_j - \langle \Delta p_j \rangle)\rangle\\
\nonumber
& &	(\Delta p_i := P_i - P_{i-1})\\
\nonumber
&=& \lim_{n \to \infty}{1 \over n} \sum_{i=1}^n \sum_{j=1-i}^{n-i}
	\langle (\Delta p_i - \langle \Delta p_i \rangle)
	(\Delta p_{i+j} - \langle \Delta p_{i+j} \rangle)\rangle\\
\nonumber
&=& \lim_{n \to \infty}{1 \over n}\sum_{i=1}^n \sum_{j=1-i}^{n-i}
	\left({A \over \alpha}\right)^2
	 \langle ( q_0 - \langle q_0\rangle )^2 \rangle C_q(j)\\
\nonumber
&=& \left({A \over \alpha}\right)^2 \lim_{n \to \infty}
	\sum_{j=1-n}^{n-1}\left( 1-{j \over n}\right)
	\langle ( q_0 - \langle q_0\rangle )^2 \rangle C_q(j)\\
&=& \left({A \over \alpha}\right)^2
\langle ( q_0 - \langle q_0\rangle )^2 \rangle
\sum_{n=-\infty}^{\infty} C_q(n),
\label{diff1}
\end{eqnarray}
if the above sum converges, i.e. if the correlations decay not slower
than $n^{-2}$.

\begin{figure}
\resizebox{\textwidth}{!}{\includegraphics{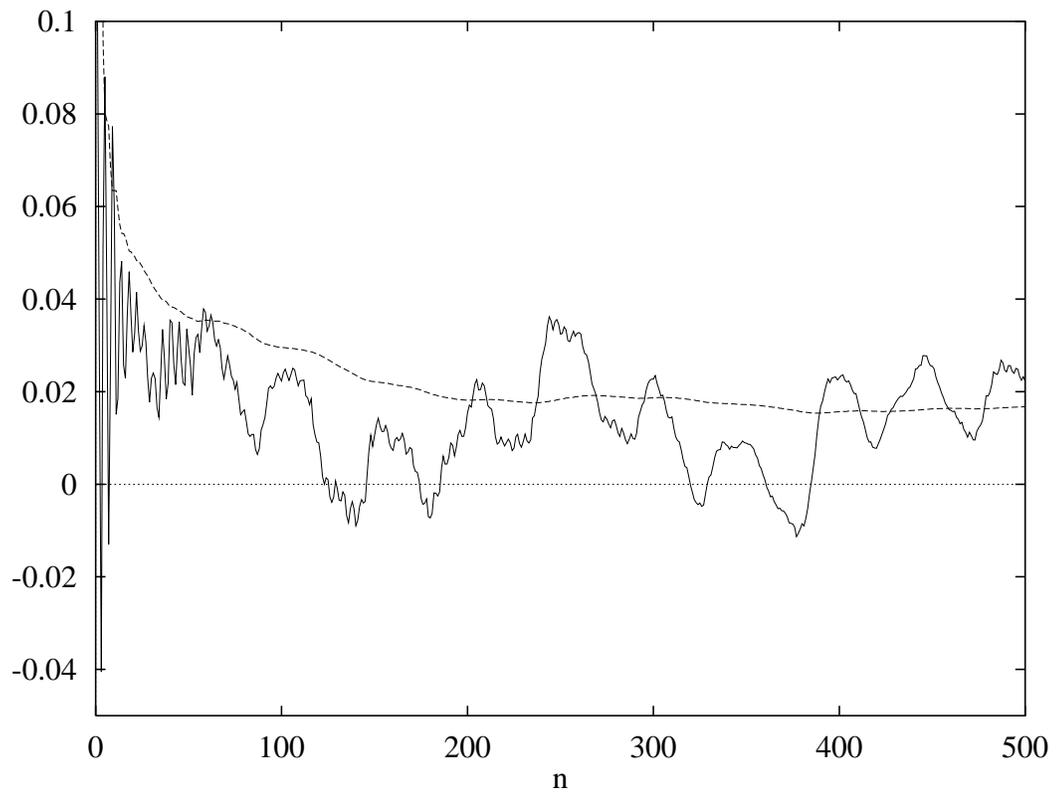}}
\caption{The comparison of the diffusion constant (dotted line)
and the sum of autocorrelation function (solid line)
in equation (\ref{diff1}) for the irrational map $M_{\gamma^2}$.}
\label{fig:diff1}
\end{figure}

\begin{figure}
\setlength{\unitlength}{1in}
\resizebox{\textwidth}{!}{
\begin{picture}(7,5)(0,0)
\put(0.05,2.5){\rotatebox{90}{\scalebox{1.5}{$\sigma^2$}}}
\put(3.8,0){\scalebox{1.5}{n}}
\put(0,0.2){\resizebox{7in}{!}{\includegraphics{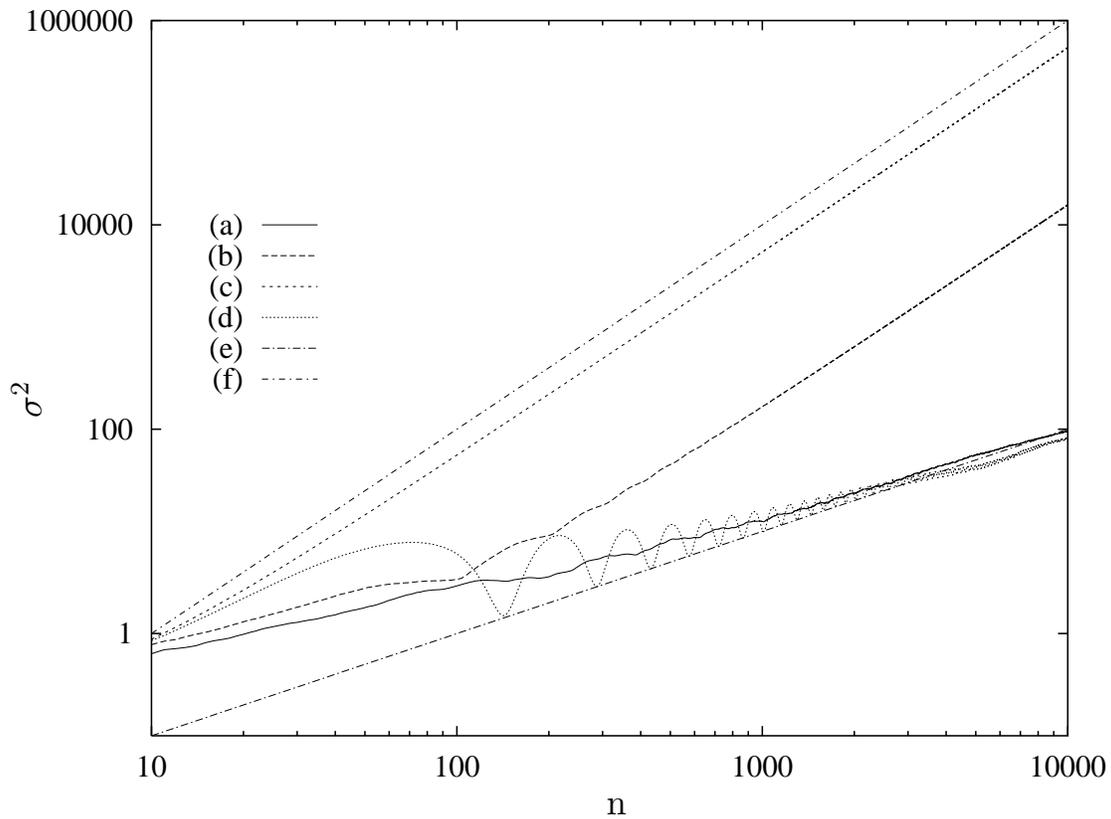}}}
\end{picture}
}
\caption{Diffusion in parabolic maps on the torus.
(a) golden map $M_{\gamma^2}$
(b) approximation to $M_{\gamma^2}$ with $\gamma = 3/5$
(c) symmetric map $M_s$
(d) approximation to $M_s$ with $\alpha = 1 + {\protect\sqrt{2} \over 100} $
(e) slope 1 (f)slope 2.
The variance $\sigma^2$ grows linearly in time for the
 irrational map $M_{\gamma^2}$ and an irrational approximation of
the symmetric map $M_s$, while for rational maps its dependence is
quadratic.}
\label{fig:diff2}
\end{figure}

To study the the diffusion rate, we took
$10^5$ initial
points distributed according to the natural measure and iterated them
$n$ times by a parabolic map $M$, and computed
$D_p(n):= \langle (P - \langle P \rangle)^2 \rangle/n$.
Figure~\ref{fig:diff1} is the comparison of this
and the sum of autocorrelation in (\ref{diff1}) performed for an
irrational map. Figure~\ref{fig:diff2} shows
the variance $\sigma^2:= \langle (P - \langle P \rangle)^2 \rangle$
 as a function of time for four different parabolic maps.
In each case, the mean was taken over $10^5$ initial points. Observe two
different kinds of behavior: for irrational maps,
the variance is proportional to time, $\sigma^2\sim D_p n$,
which corresponds to the {\sl normal diffusion}. On the other
hand, for rational maps we receive  $\sigma^2\sim D_2 n^2$, which
is called  {\sl ballistic diffusion}.
 In this case, which is sometimes referred to as the {\sl anomalous diffusion},
the relation (\ref{diff1}) looses its sense:
the right hand side of this equation does not converge and
the limit in the definition of diffusion coefficient
(\ref{diff2}) does not exist. Examples of anomalous diffusion
were reported long
time ago for a particular choice of the parameters of the standard map
\cite{Ka83}, and more recently, for the kicked Harper model
\cite{Le98}.

\section{Concluding remarks}

Although properties of linear area-preserving parabolic maps on the
plane seem to be well understood, considering the same maps on
the torus (by imposing periodic boundary conditions)
makes the behavior of the system more complicated. Generically, such
systems are discontinuous and non-invertible, and those properties
lead to relevant dynamical implications.

For rational maps,
a single trajectory is contained in a finite number of lines on the
torus. This leads to the anomalous diffusion, if the associated map 
is considered on the plane.
On the other hand, for a generic, irrational map, a typical single
trajectory explores the entire maximal invariant set. The dynamics shows
sensitivity on the initial conditions and the diffusion process is
normal.

It is worth noting that several questions concerning the parabolic maps
remain open. In particular, it would be interesting to prove the
conjecture posed in this work: the volume of the maximal invariant set
is a nowhere continuous function of the system parameter $\alpha$; it
is positive for all rational values of this parameter and zero for any
irrational $\alpha$. Moreover, our  hypothesis that the dynamics of an
irrational map is ergodic with respect to the natural invariant measure
$\mu_*$ is based on numerical experiments and still awaits a rigorous
proof.

\section{Acknowledgments}

We are indebted to P.~Ashwin for many valuable remarks and his
constant interest in the progress of this work.
We also thank M.~Arjunwadkar, U.~Feudel, C.~Grebogi, J.~Meiss,
E.~Ott, J.~Stark, M.~Wojtkowski and J.~Yorke for helpful discussions and
are
grateful to the anonymous referee for several useful comments.
K. {\.Z}. acknowledges the Fulbright Fellowship and a
support by the Polish KBN grant no. P03B~060~13.

%\newpage

{}

\newpage

\begin{figure}
\setlength{\unitlength}{1in}
\begin{picture}(6,6)(0,0)
\put(0.5,0.5){\resizebox{5in}{!}{\includegraphics{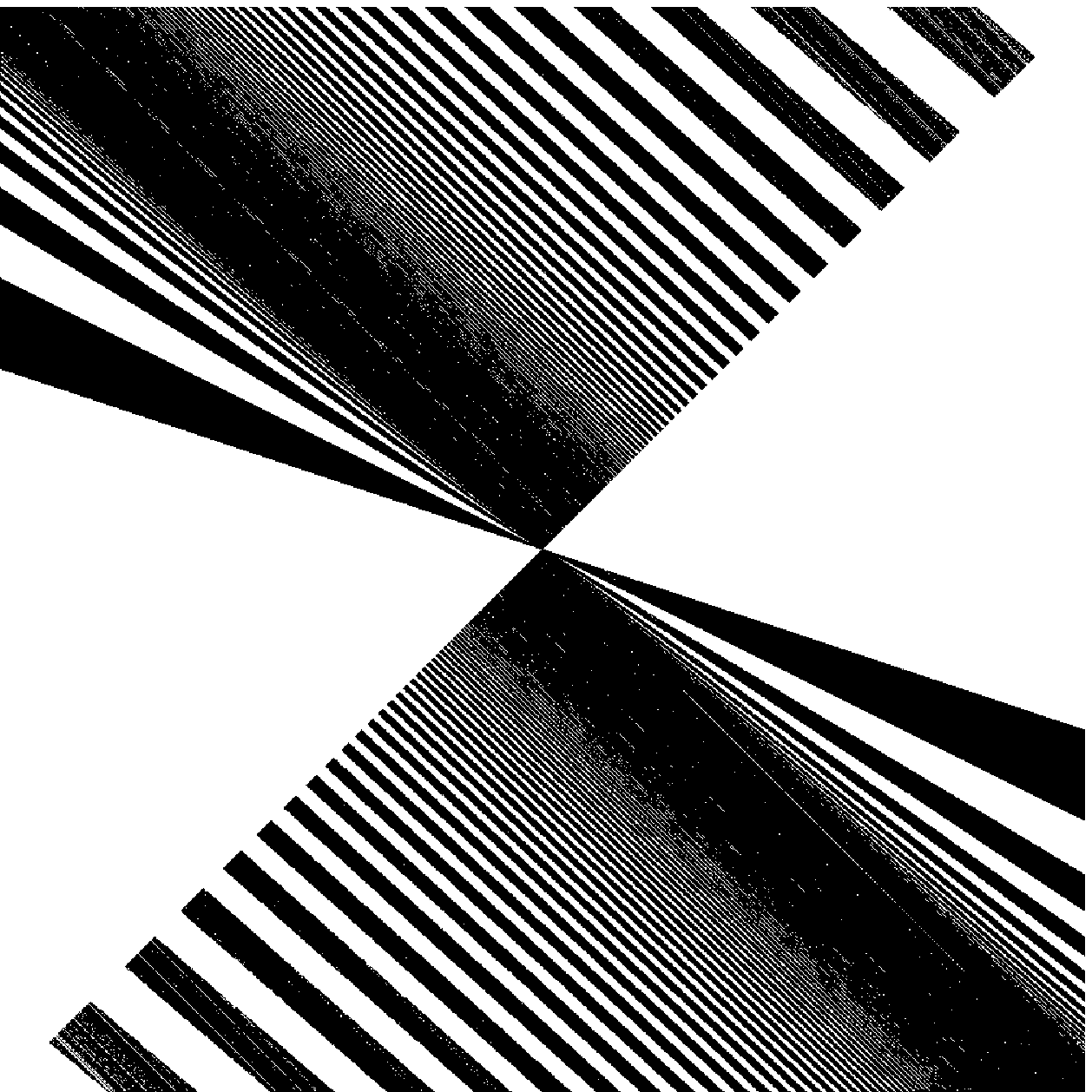}}}
\put(0.25,0.25){\Large{0.45}}
\put(0.05,5.5){\Large{0.55}}
\put(5.25,0.25){\Large{0.55}}
\put(0.5,0.5){\framebox(5,5)}
\put(0.5,5.5){\vector(0,1){0.25}}
\put(5.5,0.5){\vector(1,0){0.25}}
\put(0.45,5.85){\Large{$y$}}
\put(5.8,0.45){\Large{$x$}}
\put(3,0){\Large{(b)}}
\put(2.7,2.87){\Large{$Q$}}
\end{picture}

\end{figure}

\begin{figure}
\setlength{\unitlength}{1in}
\begin{picture}(6,6)(0,0)
\put(0.5,0.5){\resizebox{5in}{!}{\includegraphics{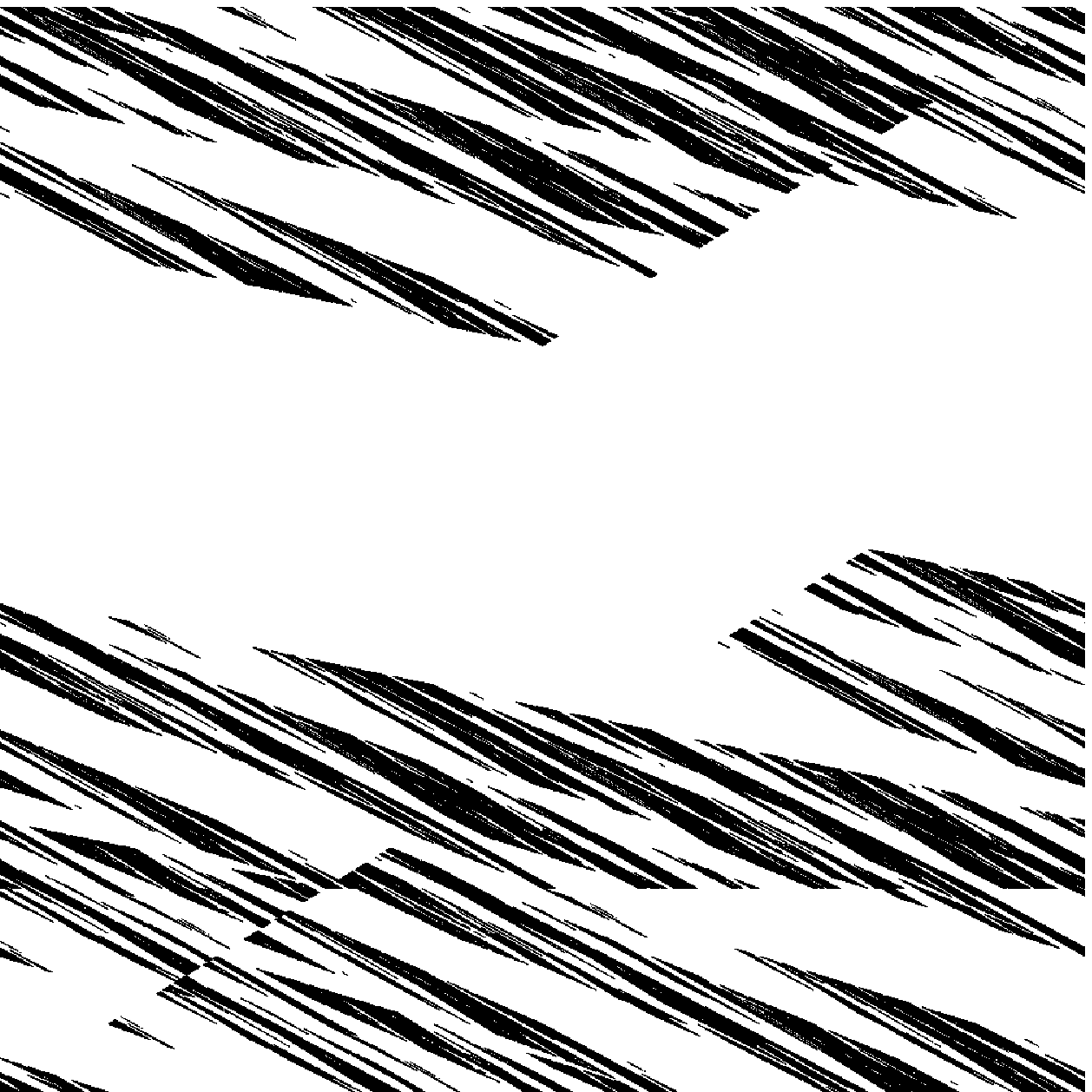}}}
\put(0.25,0.25){\Large{0}}
\put(0.25,5.5){\Large{1}}
\put(5.5,0.25){\Large{1}}
\put(0.5,0.5){\framebox(5,5)}
\put(0.5,5.5){\vector(0,1){0.25}}
\put(5.5,0.5){\vector(1,0){0.25}}
\put(0.45,5.85){\Large{$y$}}
\put(5.8,0.45){\Large{$x$}}
\put(3,0){\Large{(b)}}
\end{picture}
\end{figure}

\begin{figure}
\setlength{\unitlength}{1in}
\begin{picture}(6,6)(0,0)
\put(0.5,0.5){\resizebox{5in}{!}{\includegraphics{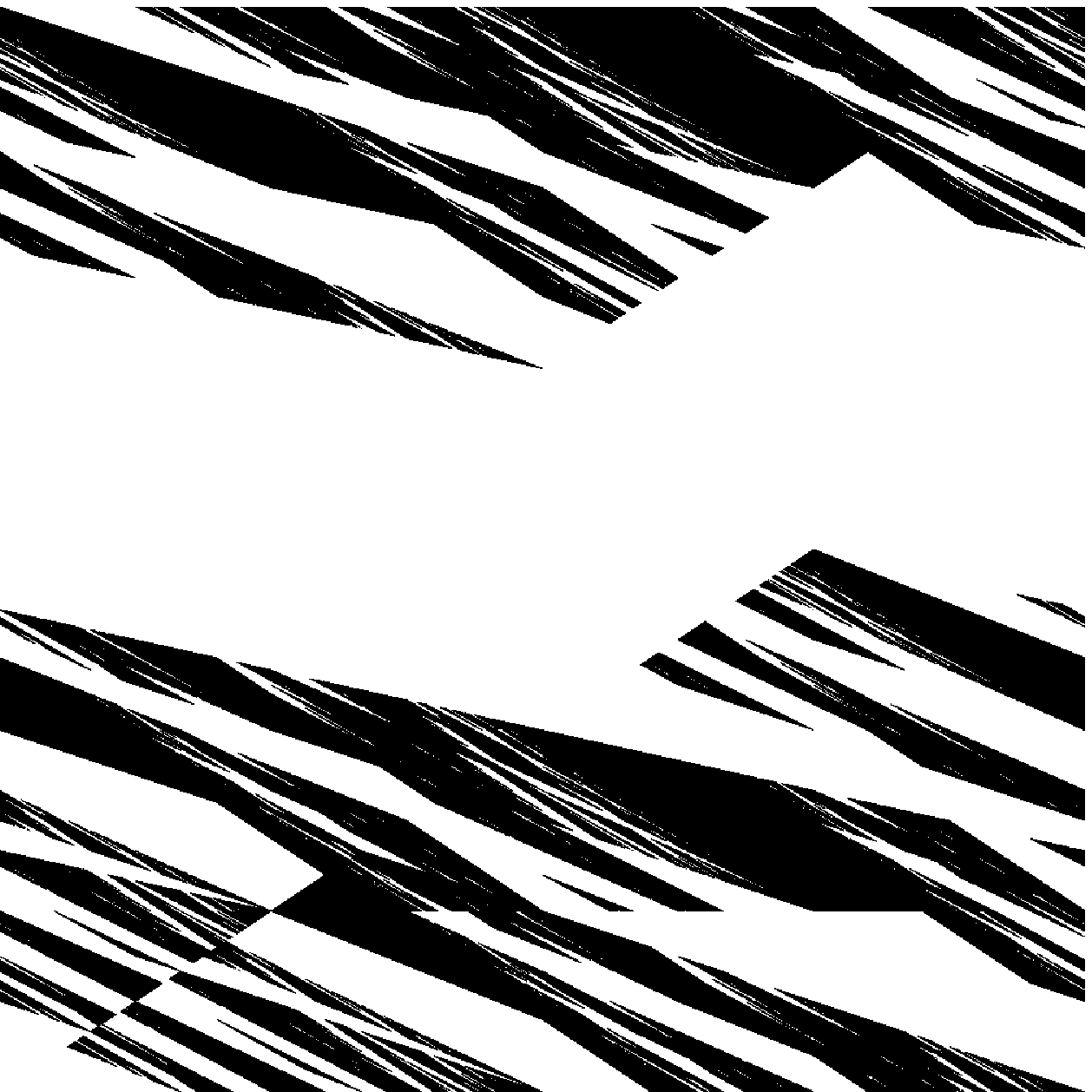}}}
\put(0.25,0.25){\Large{0}}
\put(0.25,5.5){\Large{1}}
\put(5.5,0.25){\Large{1}}
\put(0.5,0.5){\framebox(5,5)}
\put(0.5,5.5){\vector(0,1){0.25}}
\put(5.5,0.5){\vector(1,0){0.25}}
\put(0.45,5.85){\Large{$y$}}
\put(5.8,0.45){\Large{$x$}}
\put(3,0){\Large{(c)}}
\end{picture}
\end{figure}

\end{document}